\documentclass[twocolumn,preprintnumbers,amsmath,amssymb]{revtex4}

\usepackage{graphicx}
\usepackage{dcolumn}
\usepackage{bm}
\usepackage{rotating}

\usepackage{bbm}
\usepackage{verbatim}

\usepackage{color}
\definecolor{lightblue}{rgb}{0.2,0.2,0.7}
\definecolor{darkblue}{rgb}{0,0.25,0.5}
\definecolor{redbrown}{rgb}{0.875,0.25,0.125}
\definecolor{darkgreen}{rgb}{0,0.5,0}

\newcommand{\la}{\langle}
\newcommand{\ra}{\rangle}
\newcommand{\bra}[1]{\ensuremath{\langle #1 \vert}}
\newcommand{\ket}[1]{\ensuremath{\vert #1  \rangle}}

\renewcommand{\b}[1]{\ensuremath{\mathbf{#1}}}
\renewcommand{\H}{\ensuremath{\text{H}}}
\renewcommand{\l}{\ensuremath{\lambda}}

\newcommand{\lr}{\ensuremath{\text{lr}}}
\newcommand{\sr}{\ensuremath{\text{sr}}}

\newcommand{\LDA}{\ensuremath{\text{LDA}}}
\newcommand{\HF}{\ensuremath{\text{HF}}}

\newcommand{\unif}{\ensuremath{\text{unif}}}

\newcommand{\T}{\ensuremath{\text{T}}}

\DeclareMathOperator{\erf}{erf}

\begin{document}

\title{Assessment of range-separated time-dependent density-functional theory for calculating $C_6$ dispersion coefficients}

\author{Julien Toulouse$^{1,2}$}\email{julien.toulouse@upmc.fr}
\author{Elisa Rebolini$^1$}
\author{Tim Gould$^3$}
\author{John F. Dobson$^3$}
\author{Prasenjit Seal$^4$\footnote{Present address: Department of Chemistry and Supercomputing Institute, University of Minnesota, Minneapolis, Minnesota 55455-0431, USA}}
\author{J\'anos G. \'Angy\'an$^{4}$}\email{janos.angyan@univ-lorraine.fr}
\affiliation{
$^1$Laboratoire de Chimie Th\'eorique, Universit\'e Pierre et Marie Curie and CNRS, 75252 Paris, France\\
$^2$Laboratoire de Chimie et Physique Quantiques, IRSAMC, Universit\'e de Toulouse and CNRS, 31062 Toulouse, France\\
$^3$Queensland Micro and Nano Technology Centre, Griffith University, Nathan, Queensland, Australia\\
$^4$CRM2, Institut Jean Barriol, Universit\'e de Lorraine and CNRS, 54506 Vandoeuvre-l\`{e}s-Nancy, France
}


\date{\today}
\begin{abstract}
We assess a variant of linear-response range-separated time-dependent density-functional theory (TDDFT), combining a long-range Hartree-Fock (HF) exchange kernel with a short-range adiabatic exchange-correlation kernel in the local-density approximation (LDA) for calculating isotropic $C_6$ dispersion coefficients of homodimers of a number of closed-shell atoms and small molecules. This range-separated TDDFT tends to give underestimated $C_6$ coefficients of small molecules with a mean absolute percentage error of about 5\%, a slight improvement over standard TDDFT in the adiabatic LDA which tends to overestimate them with a mean absolute percentage error of 8\%, but close to time-dependent Hartree-Fock which has a mean absolute percentage error of about 6\%. These results thus show that introduction of long-range HF exchange in TDDFT has a small but beneficial impact on the values of $C_6$ coefficients. It also confirms that the present variant of range-separated TDDFT is a reasonably accurate method even using only a LDA-type density functional and without adding an explicit treatment of long-range correlation.
\end{abstract}

\maketitle

\section{Introduction}

It is well known that the leading term in the expansion of the London dispersion attractive interaction energy between a pair of atoms or molecules at long distance $R$ takes the form $- C_6/R^6$~\cite{Lon-ZP-30}. The $C_6$ dispersion coefficients are conveniently expressed by the Casimir-Polder formula~\cite{CasPol-PR-48,Lon-DFS-65} involving imaginary-frequency dynamic dipole polarizabilities, and can be efficiently calculated from linear-response time-dependent density-functional theory (TDDFT)~\cite{GisSniBae-JCP-95}. In such TDDFT calculations of $C_6$ coefficients, a number of approximations have been used for the Kohn-Sham exchange-correlation potential $v_{xc}$ and the corresponding response kernel $f_{xc}$, including the local-density approximation (LDA)~\cite{GisSniBae-JCP-95,MarCasMalMulBot-JCP-07,BanChaGha-JCP-07,BanAutCha-PRA-08}, generalized-gradient approximations (GGA)~\cite{OsiGisSniBae-JCP-97,OliBotMar-PCCP-11}, hybrid approximations~\cite{NorJieSer-JCP-03,JieNorSer-JCP-05,JieNorSer-JCP-06,SulNorSau-MP-12,KauNorSai-JCP-13} and optimized effective potential (OEP) approaches~\cite{ChuDal-JCP-04,Hir-JCP-05,ShiHirHir-PRA-06,HelBar-PRB-08,HelBar-JCP-10,Gou-JCP-12}. Using the generalized Casimir-Polder formula~\cite{Lon-DFS-65}, non-expanded dispersion energies can also be calculated from TDDFT~\cite{HesJan-CPL-03,MisJezSza-PRL-03}. The best results are obtained with LDA or GGA density functionals with asymptotically corrected potentials, hybrid approximations, and OEP approaches, with a typical accuracy on the $C_6$ coefficients of atoms and small molecules of the order of 5\%.

In the last decade, hybrid TDDFT approaches based on a range separation of electron-electron interactions have been increasingly used. The range-separated TDDFT approach that was first developed is based on the long-range correction (LC) scheme~\cite{TawTsuYanYanHir-JCP-04}, which combines long-range Hartree-Fock (HF) exchange with a short-range exchange density functional and a standard full-range correlation density functional. It has been demonstrated that the LC scheme corrects the underestimation of Rydberg excitation energies of small molecules~\cite{TawTsuYanYanHir-JCP-04} and the overestimation of (hyper)polarizabilities of long conjugated molecules~\cite{IikTsuYanHir-JCP-01,KamSekTsuHir-JCP-05,SekMaeKam-MP-05,SekMaeKamHir-JCP-07,JacPerScaFriKobAda-JCP-07,JacPerCioAda-JCC-08,KirBonRamChaMatSek-JCP-08,SonWatSekHir-JCP-08} usually obtained with standard (semi)local density-functional approximations. A variety of other similar range-separated TDDFT schemes have also been employed, which for example use an empirically modified correlation density functional depending on the range-separation parameter~\cite{LivBae-PCCP-07}, or introduce a fraction of HF exchange at shorter range as well~\cite{YanTewHan-CPL-04,SonTokSatWatHir-JCP-07,ChaHea-JCP-08,LanRohHer-JPCB-08,RohHer-JCP-08,AkiTen-CPL-08,SonWatNakHir-JCP-08,RohMarHer-JCP-09,AkiTen-IJQC-09,SonWatHir-JCP-09,PevTru-JPCL-11,NguDayPac-JCP-11,LinTsaLiCha-JCP-12}, such as in the CAM-B3LYP approximation~\cite{YanTewHan-CPL-04}.

Recently, some of us have studied a new variant of range-separated TDDFT~\cite{RebSavTou-JJJ-XX} based on the range-separated hybrid (RSH) scheme~\cite{AngGerSavTou-PRA-05}, which differs from the LC scheme in that it uses a short-range correlation density functional instead of a full-range one. This range-separated TDDFT approach, referred to as TDRSH, is motivated by the fact that, as for exchange, the long-range part of standard correlation density-functional approximations such as the LDA is usually inaccurate~\cite{TouColSav-PRA-04,TouColSav-JCP-05,TouColSav-MP-05}, so one may as well remove it. The TDRSH method can then be viewed as a first-level approximation before adding more accurate long-range correlation, e.g., by linear-response density-matrix functional theory (DMFT)~\cite{Per-JCP-12} or linear-response multiconfiguration self-consistent field (MCSCF) theory~\cite{FroKneJen-JCP-13}. Applied with a short-range adiabatic LDA exchange-correlation kernel, it was found that this TDRSH method gives in fact electronic excitation energies and oscillator strengths of small molecules very similar to the ones obtained by the range-separated TDDFT method based on the LC scheme, suggesting that the TDRSH method is already a reasonably accurate method even before adding explicit long-range correlations~\cite{RebSavTou-JJJ-XX}.

In this work, we further assess the TDRSH method by calculating isotropic $C_6$ dispersion coefficients of a set of closed-shell atoms and molecules. In particular, we investigate the impact of long-range HF exchange on these $C_6$ coefficients. To the best of our knowledge, the only range-separated TDDFT method that had been applied so far to the calculation of van der Waals dispersion coefficients was the one based on CAM-B3LYP~\cite{ByrCotMon-JCP-11,Dwy-THESIS-11,SulNorSau-MP-12,KauNorSai-JCP-13}, but the different results were inconclusive on whether or not long-range HF exchange brings any improvement. Hartree atomic units (a.u.) are used throughout the paper.

\section{Theory}

The isotropic $C_6$ dispersion coefficient between two subsystems $A$ and $B$ is given by the Casimir-Polder formula~\cite{CasPol-PR-48,Lon-DFS-65} (see Appendix~\ref{app:casimirpolder})
\begin{eqnarray}
C_6 = \frac{3}{\pi} \int_0^{\infty}du \, \bar{\alpha}_{A}(iu) \bar{\alpha}_{B}(iu),
\label{C6}
\end{eqnarray}
where $\bar{\alpha}_{S}(iu)=(\alpha_{S,xx}(iu)+\alpha_{S,yy}(iu)+\alpha_{S,zz}(iu))/3$ is the average imaginary-frequency dynamic dipole polarizability of subsytem $S$, which has the general expression
\begin{eqnarray}
\bar{\alpha}(iu) = \sum_n \frac{f_n}{\omega_n^2 + u^2},
\label{alphabar}
\end{eqnarray}
where the sum is over all excited states $n$, and $f_n$ and $\omega_n$ are the dipole oscillator strength and the excitation energy for the transition to the excited state $n$. 

In spin-restricted closed-shell TDDFT calculations, only singlet $\to$ singlet excitations contribute to Eq.~(\ref{alphabar}), since the singlet $\to$ triplet excitations have zero oscillator strength. In the TDRSH method~\cite{RebSavTou-JJJ-XX}, the singlet excitation energies ${^1}\omega_n$ are calculated in the basis of real-valued spatial RSH orbitals $\{ \phi_k(\b{r})\}$ from the familiar symmetric eigenvalue equation~\cite{Cas-INC-95}
\begin{eqnarray} 
{^1}\b{M} \, {^1}\b{Z}_{n} = {^1}\omega_{n}^2 \, {^1}\b{Z}_{n},
\label{MZ}
\end{eqnarray}
where ${^1}\b{Z}_{n}$ are normalized eigenvectors and ${^1}\b{M}=\left( {^1}\b{A}-{^1}\b{B} \right)^{1/2} \left( {^1}\b{A}+{^1}\b{B} \right) \left( {^1}\b{A}-{^1}\b{B} \right)^{1/2}$. The elements of the symmetric matrices ${^1}\b{A}$ and ${^1}\b{B}$ are
\begin{eqnarray}
{^1}A_{ia,jb}  &=& (\varepsilon_a - \varepsilon_i) \delta_{ij} \delta_{ab} + 2 \la aj| \hat{w}_{ee} |ib \ra  - \la aj|\hat{w}_{ee}^{\lr}|bi \ra 
\nonumber\\
&& + 2 \la aj |\,^1\hat{f}_{xc}^{\sr} |ib \ra,
\end{eqnarray}
\begin{eqnarray}
{^1}B_{ia,jb}  = 2 \la ab| \hat{w}_{ee} |ij \ra  - \la ab|\hat{w}_{ee}^{\lr}|ji \ra + 2 \la ab |\,^1\hat{f}_{xc}^{\sr} |ij \ra,
\end{eqnarray}
where $i,j$ and $a,b$ refer to occupied and virtual RSH spatial orbitals, respectively, $\varepsilon_k$ is the orbital eigenvalue of orbital $k$, $\la aj| \hat{w}_{ee} |ib \ra$ and $\la aj|\hat{w}_{ee}^{\lr}|bi \ra$ are two-electron integrals associated with the Coulomb interaction $w_{ee}(r)=1/r$ and the long-range interaction $w_{ee}^{\lr}(r) =\erf(\mu r)/r$, respectively, and $\la aj |{^1}\hat{f}_{xc}^{\sr} |ib \ra$ are the matrix elements of the singlet short-range adiabatic exchange-correlation kernel 
\begin{eqnarray}
\la aj |{^1}\hat{f}_{xc}^{\sr} |ib \ra &=& \int \phi_a(\b{r}_1) \phi_j(\b{r}_2) \, {^1}f_{xc}^{\sr}(\b{r}_1,\b{r}_2)
\nonumber\\
&&\times\phi_i(\b{r}_1) \phi_b(\b{r}_2) d\b{r}_1 d \b{r}_2.
\end{eqnarray}
where ${^1}f_{xc}^{\sr}(\b{r}_1,\b{r}_2) = \delta^2 E_{xc}^{\sr}[n]/\delta n(\b{r}_1) \delta n(\b{r}_2)$ is the second-order functional derivative of the short-range exchange-correlation density functional. The singlet dipole length oscillator strengths ${^1}f_n$ are obtained from the eigenvectors ${^1}\b{Z}_{n}$ with the following formula~\cite{Cas-INC-95}
\begin{equation}
{^1}f_n = \frac{4}{3} \sum_{\alpha=x,y,z} \left( \b{d}_{\alpha}^{\T} \cdot \left({^1}\b{A}-{^1}\b{B} \right)^{1/2} \cdot {^1}\b{Z}_n \right)^2,
\end{equation}
where the components of the vector $\b{d}_{\alpha}$ are $d_{\alpha,ia}=\int \phi_i(\b{r}) r_{\alpha} \phi_a(\b{r})d\b{r}$, i.e. the $\alpha$ Cartesian component of the transition dipole moment between the orbitals $i$ and $a$.

The range-separation parameter $\mu$ acts as the inverse of a smooth ``cut-off radius'' delimiting the long-range and short-range parts of the electron-electron interaction. For $\mu=0$, the method reduces to standard TDDFT (with a pure density functional and in the adiabatic approximation). For $\mu\to\infty$, the method reduces to standard time-dependent Hartree-Fock (TDHF).

To investigate the effect of range separation due to modification of the ground-state exchange-correlation potential $v_{xc}$ alone, without involving the exchange-correlation kernel $f_{xc}$, we also compute $C_6$ coefficients using bare (uncoupled) polarizabilities
\begin{eqnarray}
\bar{\alpha}^{0}(iu) = \sum_{ia} \frac{f_{ia}^{0}}{(\omega_{ia}^{0})^2 + u^2},
\label{alphabar0}
\end{eqnarray}
where the bare excitation energies are simply given by orbital energy differences, $\omega_{ia}^{0}=\varepsilon_a - \varepsilon_i$, and the bare dipole length oscillator strengths by
\begin{equation}
f_{ia}^{0} = \frac{4}{3} \omega_{ia}^{0} \sum_{\alpha=x,y,z} d_{\alpha,ia}^2.
\end{equation}

\begin{figure*}
\includegraphics[scale=0.30,angle=-90]{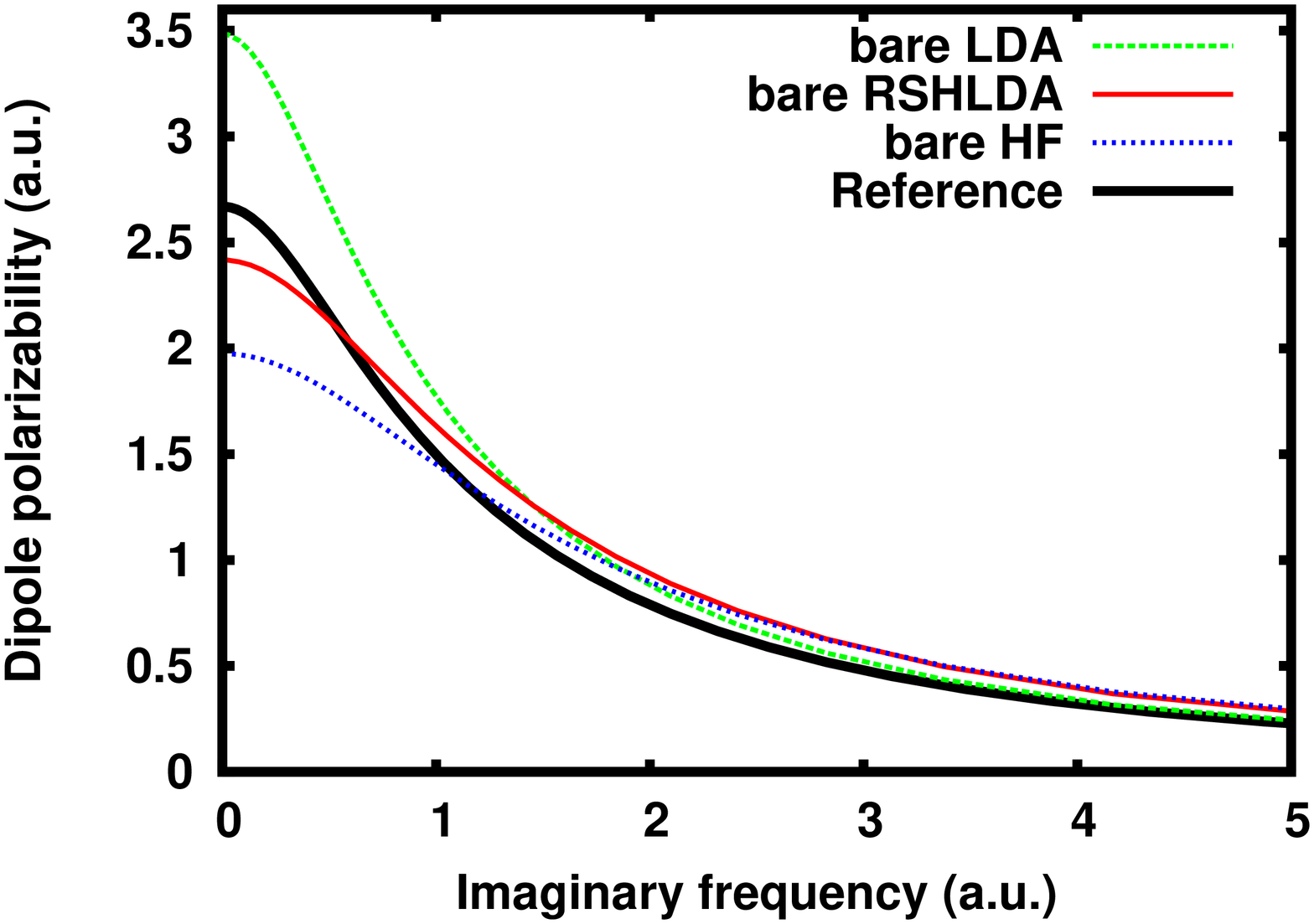}
\includegraphics[scale=0.30,angle=-90]{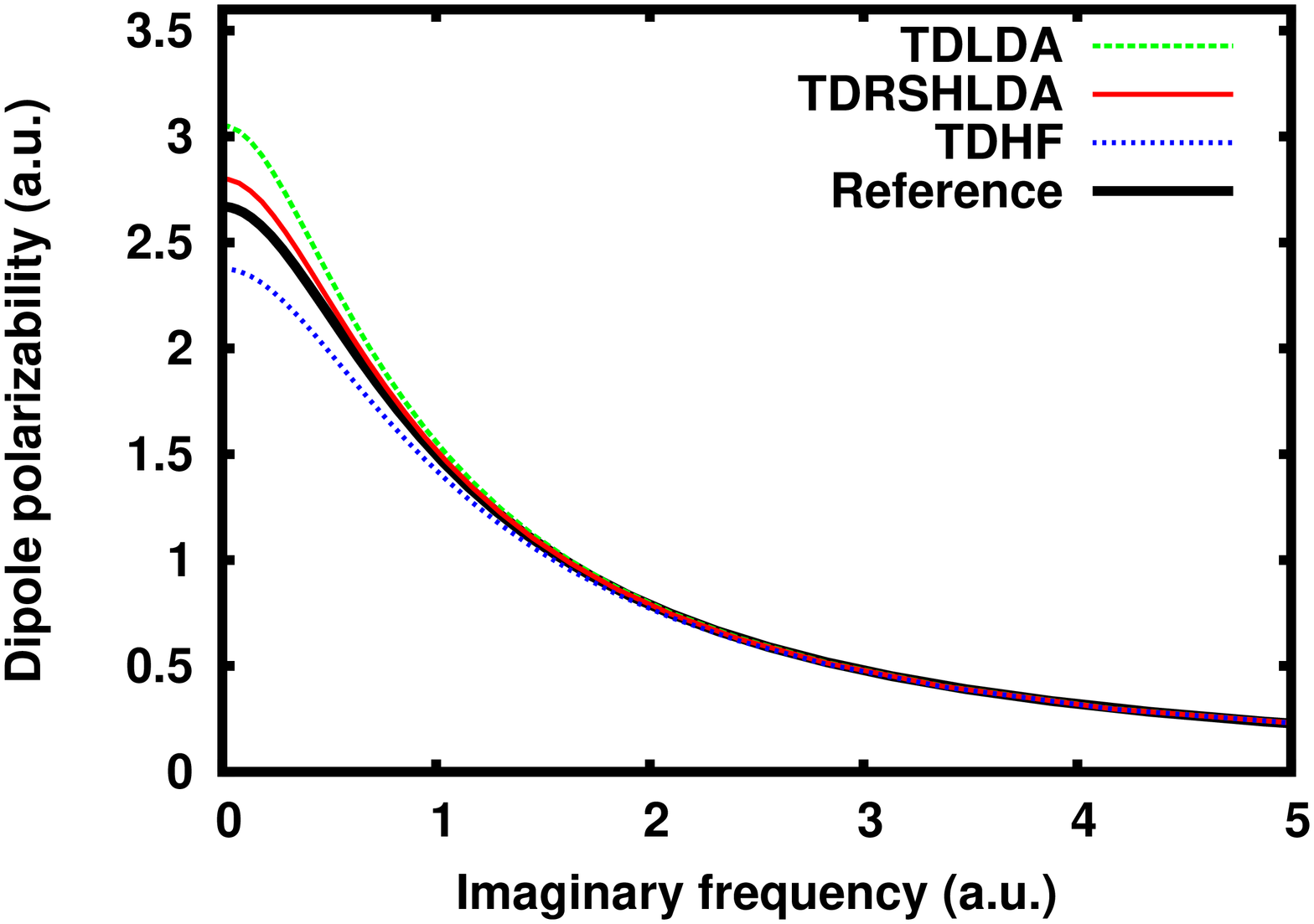}
\caption{Dynamic dipole polarizability $\bar{\alpha}(iu)$ as a function of the imaginary frequency $u$ for the Ne atom obtained by bare LDA, RSHLDA, and HF (left plot) and TDLDA, TDRSHLDA, and TDHF (right plot), with an uncontracted d-aug-cc-pCV5Z basis set. The accurate reference is taken from Ref.~\onlinecite{DerPorBab-ADNDT-10}.}
\label{fig:pola}
\end{figure*}

The exact (non-relativistic) oscillator strengths obey the well-known Thomas-Reiche-Kuhn (TRK) sum rule (or $f$-sum rule)~\cite{Tho-NAT-25,Kuh-ZP-25,ReiTho-ZP-25}
\begin{equation}
\sum_n f_{n} = N,
\end{equation}
where the sum is over all transitions and $N$ is the number of electrons. Physically, the TRK sum rule is related to the equivalence of the dipole length and dipole velocity forms of oscillator strengths, which stems from electromagnetic gauge invariance (see Ref.~\onlinecite{Fur-JCP-01}). The TRK sum rule determines the asymptotic behavior of the dynamic polarizability at large imaginary frequency, $u\to\infty$,
\begin{equation}
\bar{\alpha}(iu) \sim \frac{N}{u^2}.
\end{equation}
It has been shown that, in the limit of a complete one-electron basis set, the TRK sum rule is satisfied in TDHF~\cite{Tho-NP-61,MclBal-RMP-64,Har-JCP-69} and in TDDFT with pure density functionals (without nonlocal HF exchange)~\cite{Cas-INC-95,JamCasSal-JCP-96} or with the OEP exact-exchange approach~\cite{HelBar-PRB-08}. In the Appendix~\ref{app:TRK}, we show that the TRK sum rule is also satisfied in TDRSH, and in fact with any usual hybrid approximation, as long as the same amount of non-local HF exchange is consistently used in the ground-state potential generating the orbitals and in the response kernel. By contrast, the bare oscillator strengths in the dipole length form satisfy the TRK sum rule only if the orbitals have been generated with a local potential~\cite{Har-JCP-69}. As the HF and RSH orbitals are generated with a non-local HF exchange potential, the bare HF and RSH dipole length oscillator strengths do not sum to the number of electrons.

\section{Computational details}

The TDRSH method has been implemented for closed-shell systems in a development version of the quantum chemistry program MOLPRO~\cite{Molproshort-PROG-12}. In both the RSH ground-state potential and the response kernel, we use the short-range spin-independent (i.e., at zero spin magnetization) LDA exchange-correlation density functional
\begin{equation}
E_{xc,\LDA}^{\sr}[n] = \int n(\b{r}) \epsilon_{xc,\unif}^{\sr}(n(\b{r})) d\b{r},
\end{equation}
where $\epsilon_{xc,\unif}^{\sr}(n) = \epsilon_{xc,\unif}(n) - \epsilon_{xc,\unif}^{\lr}(n)$ is the complement short-range exchange-correlation energy per particle obtained from the exchange-correlation energy per particle of the standard uniform electron gas (UEG), $\epsilon_{xc,\unif}(n)$,~\cite{Sla-PR-51,PerWan-PRB-92} and the exchange-correlation energy per particle of a UEG with the long-range electron-electron interaction, $\epsilon_{xc,\unif}^{\lr}(n)$, as parametrized from quantum Monte Carlo calculations by Paziani \textit{et al.}~\cite{PazMorGorBac-PRB-06} (see Ref.~\onlinecite{RebSavTou-JJJ-XX} for a discussion about the corresponding kernel). For closed-shell systems, dependence on the spin magnetization needs only to be considered for triplet excitations but they do not contribute to the polarizability. The bare and response calculations are referred to as RSHLDA and TDRSHLDA, respectively. We use the value of $\mu=0.5$ bohr$^{-1}$, which was previously used in ground-state range-separated hybrid methods for applications to weak intermolecular interactions~\cite{AngGerSavTou-PRA-05,TouZhuAngSav-PRA-10,TouZhuSavJanAng-JCP-11}, without trying to re-optimize it. 

For the rare-gas and alkaline-earth-metal atoms, we use large Dunning-type uncontracted doubly-augmented core-valence quintuple-zeta quality basis sets, ensuring that the results are well converged with respect to the basis size. For He, we use the uncontracted d-aug-cc-pV5Z basis set~\cite{WooDun-JCP-94}. For all the other atoms, we have constructed uncontracted d-aug-cc-pCV5Z basis sets by augmenting available basis sets with diffuse functions using the standard even-tempered procedure. For Ne, Ar, Kr, Be, and Mg, the basis sets are obtained from the aug-cc-pCV5Z basis sets~\cite{WooDun-JCP-95,PetDun-JCP-02,DeyPetWil-JPCA-07,PraWooPetDunWil-TCA-11} by adding one diffuse function for each angular momentum of the original basis. For Ca, the basis set is obtained from the cc-pCV5Z basis set~\cite{KopPet-JPCA-02} by adding two diffuse functions for each angular momentum of the original basis. For all atoms (except, of course, He), we include all excitations from the core orbitals in the response calculation. With this setup, the TRK sum rule is very nearly satisfied, the sum of the TDLDA, TDHF, or TDRSH oscillator strengths only slightly deviating from the number of electrons by the order of $10^{-4}$ for Be, $10^{-3}$ for He, Ne, Ar, Mg, Ca, and $10^{-2}$ for Kr. While the fulfillment of the TRK sum rule to a good accuracy requires including core excitations and using very large basis sets, reasonably converged values of $C_6$ coefficients can be obtained without including core excitations and with much smaller basis sets. For example for Kr, excluding the core excitations and using the contracted d-aug-cc-pVTZ basis set gives a TDRSHLDA $C_6$ coefficient that is smaller by only about 1.5\% than the one obtained with inclusion of core excitations and with the uncontracted d-aug-cc-pCV5Z basis set. The reference values for the polarizabilities and $C_6$ coefficients of the rare-gas and alkaline-earth-metal atoms considered here are taken from Derevianko {\it et al.}~\cite{DerPorBab-ADNDT-10} and were obtained from accurate many-body calculations and/or experimental data. The contributions from relativistic effects on the value of the $C_6$ coefficients can be neglected for the atoms considered here, being at most 2\% for Ca~\cite{SulNorSau-MP-12}.

\begin{table*}
\caption{Static dipole polarizability $\bar{\alpha}(0)$ (in a.u.) for rare-gas and alkaline-earth-metal atoms obtained by bare LDA, RSHLDA, and HF, and TDLDA, TDRSHLDA, and TDHF, with uncontracted d-aug-cc-pCV5Z basis sets. 
}
\begin{tabular}{l|ccccccccc}
\hline\hline
       & bare LDA   & bare RSHLDA& bare HF     &  \phantom{xxx}  & TDLDA & TDRSHLDA& TDHF  &  \phantom{xxx}& Reference$^a$ \\
\hline                                                                                                     
He     & 1.81       & 1.24        & 1.00        &                 &  1.66 & 1.57     & 1.32  &               & 1.383         \\
Ne     & 3.48       & 2.42        & 1.98        &                 &  3.05 & 2.80     & 2.38  &               & 2.669         \\
Ar     & 18.0       & 10.8        & 10.1        &                 &  12.0 & 11.0     & 10.8  &               & 11.08         \\
Kr     & 27.8       & 16.3        & 15.9        &                 &  18.0 & 16.4     & 16.5  &               & 16.79         \\
\\                                                                                                           
Be     & 80.6       & 29.1        & 30.6        &                 &  43.8 & 43.5     & 45.6  &               & 37.76         \\
Mg     & 122        & 49.5        & 55.2        &                 &  71.4 & 73.6     & 81.6  &               & 71.26         \\
Ca     & 277        & 111         & 125         &                 &  149  & 167      & 185   &               & 157.1         \\
\hline\hline
\multicolumn{8}{l}{$^a$ From Ref.~\onlinecite{DerPorBab-ADNDT-10}.}
\end{tabular}
\label{tab:staticpola}
\end{table*}

\begin{table*}
\caption{$C_6$ coefficients (in a.u.) for homodimers of rare-gas and alkaline-earth-metal atoms obtained by bare LDA, RSHLDA, and HF, and TDLDA, TDRSHLDA, and TDHF, with uncontracted d-aug-cc-pCV5Z basis sets. 
}
\begin{tabular}{l|ccccccccc}
\hline\hline
       & bare LDA   & bare RSHLDA& bare HF    &  \phantom{xxx}& TDLDA & TDRSHLDA& TDHF  &  \phantom{xxx}& Reference$^a$ \\
\hline                                                                                     
He     & 2.17       & 1.50        & 1.12       &               &  1.86 & 1.74     & 1.37  &               & 1.461         \\
Ne     & 9.53       & 6.79        & 5.32       &               &  7.40 & 6.72     & 5.52  &               & 6.38(6)        \\
Ar     & 137        & 80.8        & 76.6       &               &  70.5 & 63.3     & 62.0  &               & 64.3(6)       \\
Kr     & 289        & 166         & 165        &               &  141  & 125      & 127   &               & 130(1)         \\
\\                                                                                                        
Be     & 642        & 232         & 255        &               &  264  & 258      & 283   &               & 214(3)        \\
Mg     & 1417       & 579         & 692        &               &  623  & 654      & 767   &               & 627(12)       \\
Ca     & 5274       & 2247        & 2693       &               & 1990  & 2374     & 2769  &               & 2121(35)      \\
\hline\hline
\multicolumn{8}{l}{$^a$ From Ref.~\onlinecite{DerPorBab-ADNDT-10}, including estimated uncertainties in parentheses.}
\end{tabular}
\label{tab:HomC6}
\end{table*}

For the molecules, we use a subset of 27 organic and inorganic molecules (going from the less polarizable H$_2$ to the most polarizable CCl$_4$) extracted from the database compiled by Tkatchenko and Scheffler~\cite{TkaSch-PRL-09}. The reference $C_6$ coefficients have been obtained from the experimental dipole oscillator strength distribution data of Meath and coworkers (see, e.g., Refs.~\onlinecite{ZeiMea-MP-77,MarMea-JCP-78}), which are believed to be accurate within 1\% - 2\%. Our $C_6$ coefficients are calculated with the d-aug-cc-pVTZ basis set~\cite{Dun-JCP-89,KenDunHar-JCP-92,WooDun-JCP-93,WilWooPetDun-JCP-99} (obtained by even-tempered augmenting the aug-cc-pVTZ basis set for Si, S, Cl, and Br) without including core excitations. The geometries were optimized with the B3LYP functional~\cite{Bec-JCP-93,BarAda-CPL-94,SteDevChaFri-JPC-94} and the aug-cc-pVDZ basis set using the quantum chemistry program GAUSSIAN~\cite{Gaussian-PROG-09}.

Since we consider relatively small systems, we can solve Eq.~(\ref{MZ}) for the full spectrum and we perform the integration over the imaginary frequency in Eq.~(\ref{C6}) analytically, giving
\begin{equation}
C_6 = \frac{3}{2} \sum_{n,m} \frac{f_{A,n} \, f_{B,m}}{\omega_{A,n} \omega_{B,m} (\omega_{A,n} + \omega_{B,m})},
\end{equation}
where $f_{S,n}$ and $\omega_{S,n}$ are the oscillator strengths and excitation energies of subsystem $S$. For large systems, the imaginary-frequency integration can done more efficiently with a numerical quadrature. 

\section{Results and discussion}

\subsection{Rare-gas and alkaline-earth-metal atoms}

As an illustrative example, we show in Fig.~\ref{fig:pola} the dynamic dipole polarizability $\bar{\alpha}(iu)$ as a function of the imaginary frequency $u$ for the Ne atom obtained by bare LDA, RSHLDA, and HF calculations and TDLDA, TDRSHLDA, and TDHF response calculations. The different methods mostly differ at small imaginary frequency. Compared to the accurate reference, for $u\lesssim 1$, the bare LDA polarizability is too large, while the bare HF polarizability is too small. The bare RSHLDA polarizability is in between the bare LDA and HF ones and closer to the reference for $u\lesssim 1$. At large imaginary frequency, all the bare polarizabilities are close to the reference curve, but it can be seen that the bare RSH and HF polarizabilities are slightly too large. This behavior can be understood from the fact that the bare RSH or HF oscillator strengths sum to a larger value than the number of electrons (11.8 and 12.9, respectively, instead of 10), contrary to the bare LDA oscillator strengths which satisfy the TRK sum rule. The TDLDA, TDRSHLDA, and TDHF polarizabilities are more accurate than their bare counterparts. At small imaginary frequency, TDLDA slightly overestimates the polarizability, TDHF slightly underestimates it, and TDRSHLDA is very close to the reference for this system. At larger imaginary frequency, $u\gtrsim 1$, TDLDA, TDHF, and TDRSHLDA all give almost exact polarizabilities, which can be understood from the fact that they all satisfy the TRK sum rule.

\begin{table*}
\caption{Isotropic $C_6$ coefficients (in a.u.) for homodimers of a subset of 27 organic and inorganic molecules extracted from the database compiled by Tkatchenko and Scheffler~\cite{TkaSch-PRL-09} obtained by bare LDA, RSHLDA, and HF, and TDLDA, TDRSHLDA, and TDHF, with d-aug-cc-VTZ basis sets. The geometry were optimized at the B3LYP/aug-cc-pVDZ level. Mean percentage errors (M\%E) and mean absolute percentage errors (MA\%E) over all molecules with respect to the reference values are given.}
\begin{tabular}{l|ccccccccc}
\hline\hline
                 & bare LDA  & bare RSHLDA & bare HF   &\phantom{xxx}& TDLDA       & TDRSHLDA  & TDHF   & \phantom{xxx}  & Reference$^a$  \\
\hline
 H$_2$           & 19.9      &    11.1     &   10.1    &             &  14.2	   &  12.7    &   12.1  &                &         12.1	\\
 HF              & 32.7      &    20.4     &   16.8    &             &  22.2	   &  19.2    &   16.7  &                &         19.0	\\
 H$_2$O          & 83.3      &    47.4     &   41.7    &             &  51.3	   &  43.4    &   40.2  &                &         45.3 	\\
 N$_2$           & 178.9     &    104.9    &   98.5    &             &  77.8	   &  72.7    &   73.7  &                &         73.3 	\\
 CO              & 182.1     &    101.8    &   91.6    &             &  84.7	   &  77.1    &   75.2  &                &         81.4     \\
 NH$_3$          & 164.8     &    89.2     &   82.5    &             &  95.9	   &  80.8    &   78.8  &                &         89.0 	\\
 CH$_4$          & 239.6     &    132.2    &   122.6   &             &  136.0	   &  121.2   &   120.4 &                &         129.7 	\\
 HCl             & 294.2     &    158.6    &   154.0   &             &  139.1	   &  122.9   &   123.7 &                &         130.4 	\\
 CO$_2$          & 391.7     &    211.4    &   179.6   &             &  163.1	   &  150.9   &   143.4 &                &         158.7 	\\
 H$_2$CO         & 312.6     &    167.8    &   150.6   &             &  155.7	   &  138.4   &   136.3 &                &         165.2 	\\
 N$_2$O          & 580.5     &    295.7    &   255.3   &             &  189.9	   &  179.8   &   177.0 &                &         184.9 	\\
 C$_2$H$_2$      & 494.8     &    260.8    &   263.1   &             &  217.9	   &  198.9   &   214.8 &                &         204.1 	\\
 HBr             & 517.2     &    269.7    &   270.3   &             &  232.9	   &  205.5   &   212.1 &                &         216.6 	\\
 H$_2$S          & 540.3     &    269.3    &   263.1   &             &  237.8	   &  209.0   &   214.1 &                &         216.8 	\\
 CH$_3$OH        & 422.9     &    231.6    &   207.9   &             &  234.0	   &  205.0   &   199.9 &                &         222.0 	\\
 SO$_2$          & 958.8     &    461.8    &   399.1   &             &  325.6	   &  295.3   &   288.4 &                &         294.0 	\\
 C$_2$H$_4$      & 645.9     &    342.3    &   333.5   &             &  313.8	   &  287.3   &   303.8 &                &         300.2 	\\
 CH$_3$NH$_2$    & 595.1     &    319.1    &   294.2   &             &  321.6	   &  279.6   &   277.9 &                &         303.8 	\\
 SiH$_4$         & 767.1     &    344.5    &   310.3   &             &  382.4	   &  329.6   &   319.3 &                &         343.9 	\\
 C$_2$H$_6$      & 742.6     &    401.3    &   372.2   &             &  395.9	   &  352.7   &   353.5 &                &         381.9 	\\
 Cl$_2$          & 1092.4    &    551.4    &   527.1   &             &  420.8	   &  385.4   &   395.7 &                &         389.2 	\\
 CH$_3$CHO       & 916.6     &    470.0    &   423.3   &             &  444.5	   &  386.6   &   381.3 &                &         401.7 	\\
 COS             & 1410.9    &    671.6    &   617.5   &             &  453.6	   &  425.4   &   429.7 &                &         402.2 	\\
 CH$_3$OCH$_3$   & 1079.7    &    570.0    &   512.2   &             &  571.9	   &  496.1   &   488.3 &                &         534.1 	\\
 C$_3$H$_6$      & 1447.4    &    746.5    &   710.6   &             &  693.6	   &  622.0   &   643.8 &                &         662.1 	\\
 CS$_2$          & 3745.6    & 	 1604.3    &   1538.1  &             &  967.0	   &  923.0   &   962.7 &                &         871.1 	\\
 CCl$_4$         & 5893.6    & 	 2792.4    &   2642.9  &             & 2186.7	   &  1924.9  &  1956.5 &                &         2024.1 	\\
\\                                                                                                        
M\%E             & 137\%     & 22.9\%      & 13.5\%    &             &  7.6\%      & -3.8\%   & -4.4\% &                &               \\
MA\%E            & 137\%     & 23.6\%      & 19.6\%    &             &  8.0\%      &  5.2\%   &  6.3\% &                &               \\
\hline\hline
\multicolumn{8}{l}{$^a$ From Ref.~\onlinecite{TkaSch-PRL-09}, obtained from experimental dipole oscillator strength distribution data.}
\end{tabular}
\label{tab:HomC6molecules}
\end{table*}

Table~\ref{tab:staticpola} reports static dipole polarizabilities $\bar{\alpha}(0)$ for rare-gas and alkaline-earth-metal atoms obtained by bare and response calculations. Bare LDA always greatly overestimates the static polarizabilities, while bare RSHLDA and HF underestimate them.TDLDA, TDRSHLDA, and TDHF give overall more accurate static polarizabilities than the bare calculations. While TDLDA decreases static polarizabilities in comparison to bare LDA, an effect that is often understood as the screening of the perturbed potential due to the response of the Hartree-exchange-correlation potential, we note that TDHF increases static polarizabilities in comparison to bare HF. Different trends are observed for the effect of HF exchange in the rare-gas atoms and in the alkaline-earth-metal atoms. For He, Ne, Ar, and Kr, starting from TDLDA which systematically overestimates the static polarizabilities, the increase of the amount of HF exchange with TDRSHLDA decreases the polarizabilities, eventually leading to a systematic underestimation in TDHF. This is consistent with the well-known tendency of TDLDA to underestimate Rydberg excitation energies and that of TDHF to overestimate them. For Be, Mg, and Ca, increasing the amount of HF exchange leads to the increase of the static polarizabilities, with TDHF systematically overestimating them. For these systems, TDHF can indeed be expected to underestimate the low-lying singlet valence excitation energy due to the fact that HF misses the important $s$-$p$ near-degeneracy ground-state correlation effects.

Table~\ref{tab:HomC6} reports $C_6$ coefficients for homodimers of rare-gas and alkaline-earth-metal atoms. As for static polarizabilities, bare LDA largely overestimates the $C_6$ coefficients for all atoms, often by more than a factor of 2, as already known~\cite{GorHeiLev-JMS-00}. Bare RSHLDA and bare HF overestimate on average the $C_6$ coefficients, whereas they underestimate static polarizabilities, meaning that they must overestimate dynamic polarizabilities at larger imaginary frequencies (as shown in Fig.~\ref{fig:pola}). TDLDA, TDRSHLDA, and TDHF give on average more accurate $C_6$ coefficients than the bare calculations. For He, Ne, Ar, and Kr, TDLDA systematically overestimates the $C_6$ coefficients, and TDRSHLDA and TDHF perform better. For Be, Mg, and Ca, TDHF greatly overestimates the $C_6$ coefficients, and TDLDA and TDRSHLDA are more accurate.

\subsection{Molecules}

Table~\ref{tab:HomC6molecules} reports isotropic $C_6$ coefficients for homodimers of 27 organic and inorganic small molecules. Mean percentage errors (M\%E) and mean absolute percentage errors (MA\%E) over all molecules with respect to the reference values are given. Overall, the same trends than those found for the rare-gas atoms are observed for these molecules. Bare LDA largely overestimates the $C_6$ coefficients, by as much as 137\%. Bare RSHLDA and HF overestimate them on average with a MA\%E of about 20\%. TDLDA, TDRSHLDA, and TDHF give $C_6$ coefficients with overall comparable accuracy, TDRSHLDA having a slightly smaller MA\%E of 5.2\% in comparison to the MA\%Es of TDLDA and TDHF, 8.0\% and 6.3\%, respectively. As for the rare-gas atoms, TDLDA overestimates the $C_6$ coefficients (with the only exception of H$_2$CO), and TDRSHLDA and TDHF give smaller $C_6$ coefficients which tend to be underestimated.

It is interesting to discuss the present results in relation with supermolecular methods which aim at describing dispersion interactions at all intermolecular distances $R$ in a seamless manner. It is well-known that the long-distance expansion of the second-order M{\o}ller-Plesset (MP2) correlation energy (using HF orbitals) gives a leading term $-C_6/R^6$ with a bare HF $C_6$ coefficient~\cite{SzaOst-JCP-77}. Similarly, the long-distance expansion of the range-separated MP2 method of Ref.~\onlinecite{AngGerSavTou-PRA-05} gives a leading term $-C_6/R^6$ with a bare RSH $C_6$ coefficient. Thus Table~\ref{tab:HomC6molecules} shows that both standard MP2 and range-separated MP2 (with the short-range LDA density functional) overestimate dispersion interactions by about 20\% at long distances for the molecules considered here,  which stresses the need to go beyond second-order perturbation theory. We note in passing that our results confirm that a supermolecular MP2 calculation using LDA orbitals (which corresponds to bare LDA $C_6$ coefficients) largely overestimates dispersion interactions at long distances~\cite{EngBon-IJMPB-01,Eng-INC-03}. Szabo and Ostlund~\cite{SzaOst-JCP-77} have found a correlation energy expression based on a variant of the random-phase approximation (RPA) (with exchange terms) which exactly gives TDHF $C_6$ coefficients in the long-distance expansion. A range-separated version of this RPA variant was found to give quite accurate dispersion interaction energies of molecular dimers around equilibrium distances~\cite{TouZhuSavJanAng-JCP-11} but the corresponding $C_6$ coefficients tend to be less accurate than in TDHF. Since we have seen that TDRSHLDA gives relatively good $C_6$ coefficients, one could try to develop a range-separated RPA-type method that still performs well at equilibrium distances and gives TDRSHLDA $C_6$ coefficients in the long-distance limit.

\section{Conclusion}

We have tested a variant of linear-response range-separated TDDFT, referred to as TDRSHLDA, combining a long-range HF exchange kernel with a short-range adiabatic LDA exchange-correlation kernel for calculating isotropic $C_6$ dispersion coefficients of homodimers of a number of closed-shell atoms and small molecules. TDRSHLDA gives $C_6$ coefficients of small molecules with a mean absolute percentage error of 5.2\%, a slight improvement over TDLDA which has a mean absolute percentage error of 8.0\%, but close to TDHF which has a mean absolute percentage error of 6.3\%. In comparison to standard TDLDA which almost always overestimates the $C_6$ coefficients, introduction of long-range HF exchange gives smaller $C_6$ coefficients (with the exceptions of the Mg and Ca atoms) which tend to be underestimated.

Our results thus show that introduction of long-range HF exchange in TDDFT has a small but beneficial impact on the values of $C_6$ coefficients of closed-shell atoms and small molecules. According to previous studies on (hyper)polarizabilities~\cite{IikTsuYanHir-JCP-01,KamSekTsuHir-JCP-05,SekMaeKam-MP-05,SekMaeKamHir-JCP-07,JacPerScaFriKobAda-JCP-07,JacPerCioAda-JCC-08,KirBonRamChaMatSek-JCP-08,SonWatSekHir-JCP-08}, a bigger impact can be expected for larger molecules. More importantly, this work confirms the conclusion of a previous study on excitation energies and oscillator strengths of a few small molecules~\cite{RebSavTou-JJJ-XX} in that the TDRSH method is a reasonably accurate method even using only a LDA-type density functional and without adding an explicit treatment of long-range correlation.

\section*{Acknowledgments}
This work was partly supported by the French Australian Science and Technology (FAST) Program. JT thanks Trond Saue for discussions, and gratefully acknowledges the hospitality of the Laboratoire de Chimie et Physique Quantiques (Universit\'e de Toulouse and CNRS, Toulouse, France) where part of this research was done. JFD and TG were supported by Australian Research Council Discovery Grant DP1096240. PS was supported by Agence Nationale de la Recherche (contract 07-BLAN-0272 ``Wademecom'').

\appendix
\section{Casimir-Polder formula}
\label{app:casimirpolder}

The dispersion energy between two subsystems $A$ and $B$ in their ground states is defined by the following second-order energy correction in Rayleigh-Schr\"odinger intermolecular perturbation theory
\begin{equation}
 E_{disp}^{AB} = -\sum_{a\neq 0}\sum_{b\neq 0} \frac{\vert \bra{\Psi^A_0\Psi^B_0}\hat{W}_{AB}\ket{\Psi^A_a\Psi^B_b}\vert^2}{E^A_a+E^B_b -E^A_0-E^B_0},
\label{EdispAB2ndorder}
\end{equation}
where $\Psi^A_0$ and $\Psi^A_a$ are the ground and excited states of $A$ with associated energies $E^A_0$ and $E^A_a$, and similarly for $B$, $\hat{W}_{AB}=\iint \hat{n}_A(\b{r}_1) w_{ee}(|\b{r}_2-\b{r}_1|)\hat{n}_B(\b{r}_2)d\b{r}_1d\b{r}_2$ is the intermolecular electron-electron interaction operator written in terms of the density operators of $A$ and $B$. In Eq.~(\ref{EdispAB2ndorder}), it has been assumed that the two subsystems $A$ and $B$ are sufficiently far apart (non-overlapping) so that exchange contributions between them can be neglected. Using the integral transform, $1/(x+y)=(2/\pi) \int_0^{\infty} \left[ x/(x^2+u^2) \right] \, \left[y/(y^2+u^2)\right] \, du$ for $x,y>0$, which permits to recast the energy denominator in Eq.~(\ref{EdispAB2ndorder}) into a multiplicative separable form, and using the definition of the imaginary-frequency linear density-density response function of a subsystem in terms of its eigenstates $\Psi_k$ and excitation energies $\omega_{k}=E_k-E_0$, for real-valued matrix elements $\bra{\Psi_0} \hat{n}(\b{r})\ket{\Psi_k}$,
\begin{equation}
\chi(\b{r},\b{r}';i u)= -\sum_{k\neq 0} \frac{2\omega_{k}
    \bra{\Psi_0} \hat{n}(\b{r})\ket{\Psi_k}
          \bra{\Psi_k} \hat{n}(\b{r}')\ket{\Psi_0}}
    {\omega_{k}^2 + u^2},
\end{equation}
one easily arrives to the generalized Casimir-Polder formula~\cite{Lon-DFS-65,ZarKoh-PRB-76,DmiPei-IJQC-81,Mcw-CCA-84,McW-BOOK-92,Dob-INC-94,DobMcLRubWanGouLeDin-AJC-01}
\begin{eqnarray}
E_{disp}^{AB} &=& -\frac{1}{2\pi} \int_0^{\infty}du  \int d\b{r}_1 d\b{r}_1' d\b{r}_2 d\b{r}_2' \, \chi_A (\b{r}_1,\b{r}_1';iu)
\nonumber\\
&& w_{ee}(|\b{r}_2-\b{r}_1|) \chi_B (\b{r}_2,\b{r}_2';iu) w_{ee}(|\b{r}_2'-\b{r}_1'|),
\nonumber\\
\end{eqnarray}
where $\chi_{S} (\b{r},\b{r}';iu)$ is the linear response function of the subsystem $S=A\text{ or }B$. Assuming that the subsystems $A$ and $B$ are separated by a large vector $\b{R}$, one can perform a multipolar expansion of the Coulomb interaction $w_{ee}(|\b{r}_2-\b{r}_1|)=1/|\b{r}_2-\b{r}_1|$, redefining the origins of $\b{r}_1$ and $\b{r}_2$ at either ends of $\b{R}$,
\begin{eqnarray}
w_{ee}(|\b{r}_2-\b{r}_1|) &=& \frac{1}{R} + \sum_\alpha T_\alpha(\b{R}) (\b{r}_2-\b{r}_1)_\alpha 
\nonumber\\
 &&+ \frac{1}{2} \sum_{\alpha,\beta} T_{\alpha\beta}(\b{R}) (\b{r}_2-\b{r}_1)_\alpha (\b{r}_2-\b{r}_1)_\beta
\nonumber\\
 &&+ \cdots,
\end{eqnarray}
with the Cartesian interaction tensors $T_\alpha(\b{R})=-R_\alpha R^{-3}$ and $T_{\alpha\beta}(\b{R})=(3 R_\alpha R_\beta - R^2 \delta_{\alpha\beta}) R^{-5}$, the Greek indices referring to $x$, $y$ or $z$ components. Because the integration of $\chi(\b{r},\b{r}';i u)$ over $\b{r}$ and $\b{r}'$ is zero (normalization of the perturbed density and zero-response to a uniform perturbative potential), only the terms containing products of components of the four coordinates $\b{r}_1$, $\b{r}_1'$, $\b{r}_2$, $\b{r}_2'$ survive in the leading term of the dispersion energy,
\begin{eqnarray}
E_{disp}^{AB} = -\frac{1}{2\pi} \sum_{\alpha,\beta,\gamma,\delta} T_{\alpha\beta}(\b{R}) T_{\gamma\delta}(\b{R}) \;\;\;\;\;\; \;\;\;\;\;\;
\nonumber\\
\times\int_0^{\infty}du \, \alpha_{A,\delta\alpha}(iu) \alpha_{B,\beta\gamma}(iu) +\cdots, \;\;\;
\label{EdispABsum}
\end{eqnarray}
where $\alpha_{S,\alpha\beta}(iu) = -\int d\b{r} d\b{r}' \chi_{S} (\b{r},\b{r}';iu) \b{r}_{\alpha} \b{r}'_{\beta}$ is the $\alpha,\beta$ Cartesian component of the imaginary-frequency dynamic dipole polarizability tensor of the subsystem $S$. If we consider the spherically averaged dipole polarizability, $\alpha_{S,\alpha\beta}(iu) = \bar{\alpha}_{S}(iu) \delta_{\alpha\beta}$ where $\bar{\alpha}_{S}(iu)=(\alpha_{S,xx}(iu)+\alpha_{S,yy}(iu)+\alpha_{S,zz}(iu))/3$, then it is easy to do the sum over $\alpha$, $\beta$, $\gamma$, $\delta$ in Eq.~(\ref{EdispABsum}) to get the familiar Casimir-Polder formula for the leading term of the dispersion energy
\begin{eqnarray}
E_{disp}^{AB} = -\frac{3}{\pi R^6} \int_0^{\infty}du \, \bar{\alpha}_{A}(iu) \bar{\alpha}_{B}(iu) +\cdots,
\end{eqnarray}
where we have used that $\sum_{\alpha \beta} T_{\alpha\beta}(\b{R}) T_{\beta\alpha}(\b{R}) =6/R^6$.

\section{Thomas-Reiche-Kuhn sum rule in TDRSH}
\label{app:TRK}

In this Appendix, we show that, in the limit of a complete one-electron basis set, the TRK sum rule of oscillator strengths, $\sum_n f_n =N$ where $N$ is the number of electrons, holds in TDRSH and in fact more generally in TDDFT with any hybrid approximation including nonlocal HF exchange. It is well known that the TRK sum rule holds in TDHF~\cite{Tho-NP-61,MclBal-RMP-64,Har-JCP-69,HanBou-MP-79,HarHanBou-JCP-89}, and in TDDFT with pure density functionals~\cite{Cas-INC-95,JamCasSal-JCP-96} or with the OEP exact-exchange approach~\cite{HelBar-PRB-08}, but we have not found the explicit proof in the literature for TDDFT with hybrid approximations. The TRK sum rule must hold in this case as well, as it more generally holds for linear response on variational ground-state many-body theories~\cite{Str-RNC-88,LeeDah-INC-94}. Nevertheless, an explicit proof is interesting since it reveals that the key to the fulfillment of the TRK sum rule is to use the same amount of HF exchange in the ground-state potential and in the response kernel.

Throughout this Appendix, we work with a real-valued spin-orbital basis without spin adaptation. The oscillator strengths are~\cite{Cas-INC-95}
\begin{eqnarray}
f_n &=& \frac{2}{3} \sum_{\alpha=x,y,z} \left( \b{d}_{\alpha}^{\T} \cdot \left(\b{A}-\b{B} \right)^{1/2} \cdot \b{Z}_n \right)^2,
\end{eqnarray}
where $\b{d}_{\alpha}$ is the $\alpha$-component transition moment vector, $\b{Z}_n$ are the normalized eigenvectors of $\b{M}=\left(\b{A}-\b{B} \right)^{1/2} \left( \b{A}+\b{B} \right) \left( \b{A}-\b{B} \right)^{1/2}$, and in TDRSH the matrices $\b{A}$ and $\b{B}$ have elements
\begin{eqnarray}
A_{ia,jb}  &=& (\varepsilon_a - \varepsilon_i) \delta_{ij} \delta_{ab} + \la aj| \hat{w}_{ee}^{\lr} |ib \ra  - \la aj|\hat{w}_{ee}^{\lr}|bi \ra 
\nonumber\\
&& + \la aj |\hat{f}_{\H xc}^{\sr} |ib \ra,
\label{Aiajb}
\end{eqnarray}
and
\begin{eqnarray}
B_{ia,jb}  = \la ab| \hat{w}_{ee}^{\lr} |ij \ra  - \la ab|\hat{w}_{ee}^{\lr}|ji \ra + \la ab |\hat{f}_{\H xc}^{\sr} |ij \ra,
\label{Biajb}
\end{eqnarray}
where $i,j$ and $a,b$ refer to occupied and virtual RSH spin-orbitals, respectively, $\varepsilon_k$ is the orbital eigenvalue of spin-orbital $k$, $\la aj| \hat{w}_{ee}^{\lr} |ib \ra$ are two-electron integrals associated with the long-range interaction, and $\la aj |\hat{f}_{\H xc}^{\sr} |ib \ra$ are the integrals associated with the short-range Hartree-exchange-correlation (Hxc) kernel. Since in the adiabatic approximation the eigenvectors $\b{Z}_n$ form a complete orthonormal basis (for any selection of single excitations and one-electron basis set), one can use the completeness relation, $\sum_n \b{Z}_n \b{Z}_n^{\T} = \b{1}$, to obtain~\cite{Cas-INC-95,JamCasSal-JCP-96}
\begin{eqnarray}
\sum_n f_n &=& \frac{2}{3} \sum_{\alpha=x,y,z} \b{d}_{\alpha}^{\T} \cdot \left(\b{A}-\b{B} \right) \cdot  \b{d}_{\alpha},
\nonumber\\
&=& \frac{2}{3} \sum_{\alpha=x,y,z} \sum_{ia,jb} d_{\alpha,ia} \left( A_{ia,jb} - B_{ia,jb} \right) d_{\alpha,jb}.
\nonumber\\
\label{sumfn}
\end{eqnarray}
In TDDFT with pure density functionals, the matrix $\b{A}-\b{B}$ is diagonal and contains the orbital energy differences, and one recovers the sum of the bare KS oscillator strengths, $\sum_n f_n = \sum_{ia} f_{ia}^{0}$, which trivially satisfies the TRK sum rule in a complete one-electron basis (and considering in the sum over $ia$ all single excitations including those from the core orbitals) due to the locality of the KS potential. When including nonlocal HF exchange, however, the matrix $\b{A}-\b{B}$ is no longer diagonal.

Using a second-quantized equations-of-motion formalism (see Ref.~\onlinecite{Row-RMP-68}), the elements of the TDRSH matrices $\b{A}$ and $\b{B}$ can be conveniently written with expectation values of double commutators over the RSH ground-state single-determinant wave function $\Phi_0$ 
\begin{eqnarray}
A_{ia,jb}  &=& \bra{\Phi_0} [ [ \hat{a}_i^\dagger \hat{a}_a, \hat{H}^\lr], \hat{a}_b^\dagger \hat{a}_j] \ket{\Phi_0}
+ \la aj |\hat{f}_{\H xc}^{\sr} |ib \ra,
\nonumber\\
\end{eqnarray}
and
\begin{eqnarray}
B_{ia,jb}  &=& -\bra{\Phi_0} [ [ \hat{a}_i^\dagger \hat{a}_a, \hat{H}^\lr], \hat{a}_j^\dagger \hat{a}_b] \ket{\Phi_0}
+ \la ab |\hat{f}_{\H xc}^{\sr} |ij \ra,
\nonumber\\
\end{eqnarray}
where the long-range effective Hamiltonian $\hat{H}^\lr = \hat{H}_0 + \hat{W}^{\lr}$ involves the RSH reference Hamiltonian $\hat{H}_0 = \hat{T} + \hat{V}_{ne} + \hat{V}_{\H x,\HF}^{\lr} + \hat{V}_{\H xc}^{\sr}$ [generating the orbital energy differences in Eq.~(\ref{Aiajb})] and the long-range fluctuation potential operator $\hat{W}^{\lr} = \hat{W}_{ee}^{\lr} - \hat{V}_{\H x,\HF}^{\lr}$ [generating the long-range two-electron integrals in Eqs.~(\ref{Aiajb}) and~(\ref{Biajb})]. In these expressions, $\hat{T}$ is the kinetic energy operator, $\hat{V}_{ne}$ is the nuclei-electron interaction operator, $\hat{V}_{\H x,\HF}^{\lr}$ is the long-range HF potential operator, $\hat{V}_{\H xc}^{\sr}$ is the short-range Hxc potential operator, and $\hat{W}_{ee}^{\lr}$ is the long-range two-electron interaction operator. The contributions from the short-range Hxc kernel $f_{\H xc}^{\sr}$ cancel out in $\b{A}-\b{B}$, and using the (second-quantized) dipole moment operator, $\hat{d}_\alpha = \sum_{kl} d_{\alpha,kl} \; \hat{a}_k^\dagger \hat{a}_l$, where each sum is over all (occupied and virtual) spin-orbitals, it can be shown that Eq.~(\ref{sumfn}) simplifies to (considering in the sums over $ia$ and $jb$ all single excitations including those from the core orbitals)
\begin{eqnarray}
\sum_n f_n  &=& \frac{2}{3} \sum_{\alpha=x,y,z} \sum_{ia} d_{\alpha,ia} \bra{\Phi_0} [ [ \hat{a}_i^\dagger \hat{a}_a, \hat{H}^\lr], \hat{d}_\alpha] \ket{\Phi_0}
\nonumber\\
&=& \frac{1}{3} \sum_{\alpha=x,y,z} \bra{\Phi_0} [ [ \hat{d}_\alpha, \hat{H}^\lr], \hat{d}_\alpha] \ket{\Phi_0},
\end{eqnarray}
where the last equality is found by taking the adjoint of the double commutator, and noticing that the diagonal terms $\hat{a}_k^\dagger \hat{a}_k$ in $\hat{d}_\alpha$ do not contribute. 

In the limit of a complete one-electron basis set, $\hat{d}_\alpha$ and all the potentials in the effective Hamiltonian $\hat{H}^\lr = \hat{T} + \hat{V}_{ne} + \hat{W}_{ee}^{\lr} + \hat{V}_{\H xc}^{\sr}$ are multiplicative local operators in the position representation and thus commute with each other, and the double commutator with the kinetic energy operator can be evaluated using the position-momentum canonical commutation relation, leading to (see, e.g., Ref.~\onlinecite{HelJorOls-BOOK-02}): $[ [ \hat{d}_\alpha, \hat{H}^\lr], \hat{d}_\alpha] = [ [ \hat{d}_\alpha, \hat{T}], \hat{d}_\alpha] = i [ \hat{p}_\alpha, \hat{d}_\alpha] = \hat{N}$ where $\hat{p}_\alpha$ is the (second-quantized) momentum operator and $\hat{N}$ is the particle-number operator. It follows that the TDRSH oscillator strengths satisfy the TRK sum rule
\begin{equation}
\sum_n f_n = N.
\end{equation}
The proof relies on the cancellation of the nonlocal HF potential $\hat{V}_{\H x,\HF}^{\lr}$ in the effective Hamiltonian $\hat{H}^\lr$, which requires that that the same amount of HF exchange is used in the ground-state potential generating the orbitals and in the response kernel. The proof can be trivially adapted to TDDFT with hybrid approximations, replacing the long-range interaction by the full-range one: $\hat{W}^{\lr} \to \l \hat{W} = \l (\hat{W}_{ee} - \hat{V}_{\H x,\HF})$ where $\l$ is the fraction of HF exchange.


\end{document}